%% file: ISWCS_2019.tex
\documentclass[conference,a4paper]{IEEEtran}

\usepackage{etex}

\def\papertitle{Early Detection for Optimal-Latency Communications in Multi-Hop Links}

\usepackage{ifpdf}
\ifpdf
\usepackage[draft,pdfborder={0 0 0}]{hyperref}
\fi

\usepackage{cite}
\usepackage{graphicx,epstopdf}

\usepackage{amssymb}
\usepackage[cmex10]{amsmath}
\usepackage[subnum]{cases}
\usepackage{nicefrac}

\newcommand{\argmax}{\operatornamewithlimits{arg\,max}}

\newtheorem{theorem}{Theorem}

\newcommand{\qed}{\hfill\IEEEQEDclosed}

\usepackage{glossaries} 
\usepackage{tikz,pgf}
\usepackage{pgfplots}

\usepackage{balance} 

\usepackage{changes}
\newcommand{\fixme}[2]{\ifx&#2&{\color{red}#1}\else{\color{red}FIXME\{}#1{\color{red}\}}\footnote{{\color{red}#2}}\PackageWarning{Fixme}{#1: #2}\fi}

\title{\papertitle}

\author{
  \IEEEauthorblockN{Diego Barrag\'an Guerrero\IEEEauthorrefmark{1}\IEEEauthorrefmark{2}, Minh Au\IEEEauthorrefmark{3}, Ghyslain Gagnon\IEEEauthorrefmark{1}, Fran\c{c}ois Gagnon\IEEEauthorrefmark{1}, and Pascal Giard\IEEEauthorrefmark{1}}
	\\
  \IEEEauthorblockA{
		\IEEEauthorrefmark{1}Electrical Engineering Department, \'Ecole de technologie sup\'erieure (\'ETS), Montr\'eal, Canada\\
		\IEEEauthorrefmark{2}Computer Science and Electronic Department, Universidad T\'ecnica Particular de Loja (UTPL), Loja, Ecuador\\
		\IEEEauthorrefmark{3}Hydro-Qu\'ebec's Research Institute (IREQ), Varennes, Canada
  }
}

\newcommand{\plotfigureheight}{0.63}

\begin{document}

\newacronym{awgn}{AWGN}{additive white Gaussian noise}
\newacronym{df}{DF}{decode-and-forward}
\newacronym{iid}{i.i.d.}{independent and identically distributed}
\newacronym{ofdm}{OFDM}{orthogonal frequency-division multiplexing}
\newacronym[plural=BLERs]{bler}{BLER}{block-error rate}
\newacronym{snr}{SNR}{signal-to-noise ratio}

\bstctlcite{IEEEexample:BSTcontrol}

\maketitle

\begin{abstract}
  Modern wireless machine-to-machine-type communications aim to provide both ultra reliability and low latency, stringent requirements that appear to be mutually exclusive.
  From the noisy channel coding theorem, we know that reliable communications mandate transmission rates that are lower than the channel capacity. To guarantee arbitrarily-low error probability, this implies the use of messages whose lengths tend to infinity. However, long messages are not suitable for low-latency communications.
  In this paper, we propose an early-detection scheme for wireless communications under a finite-blocklength regime that employs a sequential-test technique to reduce latency while maintaining reliability.
  We prove that our scheme leads to an average detection time smaller than the symbol duration.
  Furthermore, in multi-hop low-traffic or continuous-transmission links, we show that our scheme can reliably detect symbols before the end of their transmission, significantly reducing the latency, while keeping the error probability below a predefined threshold.
\end{abstract}

\section{Introduction}

Many critical applications require ultra-reliable communication \emph{and} extremely-low latency, e.g., communications between financial trading centers \cite{Karzand2015}, or vehicle-to-vehicle communications for collision warning \cite{Yang2004}. As stated by Shannon \cite{Shannon1948}, to achieve ultra-reliable communications (arbitrarily-small error probability), the blocklength must tend to infinity, inducing a high latency. Therefore, critical communication applications use channel codes where short codewords can be transmitted over a noisy channel with an average \gls{bler} smaller than a predetermined target that tends to zero.

Recent results on non-asymptotic fundamental limits of channel coding allow us to approximate the minimum blocklength needed for a targeted \gls{bler} given the capacity of a noisy channel \cite{Durisi_2016,Oestman_2017}. These fundamental results can be applied to one-hop communication systems. On the other hand, research in multi-hop networks has been devoted to constructing transmission schemes with minimal delay \cite{Wen2008,Chaaban2015,Chaaban2016,Ngo2011,Maric2013}. In these works however, the signaling delay has not been taken into account. Most of these systems need to be synchronized as they are required to wait at the end of a transmitted symbol to make a decision (i.e., \textit{synchronous detection}). Several new waveforms have been proposed to fulfill the low-latency requirements of the next-generation mobile-communication standard (5G) \cite{Fettweis2009,Wunder2014,Schaich2014}. Similarly, these waveforms all require synchronization. 

An optimal low-latency transmission strategy for schemes employing amplify-and-forward relaying was constructed \cite{Au2016}.
It used an early-detection scheme inspired by communications with sequential detection feedback where the transmission ends as soon as the receiver makes the correct decision \cite{Viterbi1965}.
However, contrary to \cite{Viterbi1965}, that early-detection scheme does not rely on a feedback.
In this paper, we investigate the minimal latency in multi-hop systems that employ \gls{df} relaying schemes, when either synchronous-detection or early-detection schemes are used.
Both work consider an \gls{ofdm}-like signal defined as simultaneous transmission of all symbols in parallel over the channel.

\subsubsection*{Contributions}
The first part of this paper presents a simplification of the early-detection technique from previous work and proves that the resulting scheme has optimal latency in a single-hop system. For simplicity, we assume throughout this paper a power-constrained memoryless \gls{awgn} channel. The second part applies this scheme to two common types of multi-hop systems with \gls{df} relays, and the latency is proved to be optimal in these systems as well.

\section{Definition and Problem Statement}
\label{sect:II}
\subsection{Definition of Channel Coding}
Consider a code with blocklength $n$ generating $M$ codewords. We denote the input and output alphabets $\mathcal{A}$ and $\mathcal{B}$, and a conditional probability measure $P_{\mathbf{Y}\mid \mathbf{X}} : \mathcal{A}^{n} \longmapsto \mathcal{B}^{n}$, in which $\mathbf{X}$ represents an input sequence encoded by an $(n,M,\epsilon)$-code, where $\epsilon>0$ is the block-error probability.
$\mathbf{Y}$, the corresponding output sequence, depends statistically on the input sequence $\mathbf{X}$ through the conditional probability density function $P_{\mathbf{Y}\mid \mathbf{X}}$. Specifically, in \gls{awgn} channels we have:
\begin{equation}
\mathbf{Y} = \mathbf{X}+\mathbf{Z}\,,
\end{equation}
where $\mathbf{X} \in \mathcal{A}^{n}$, $\mathbf{Y} \in \mathcal{B}^{n}$ and $\mathbf{Z}$ is a random vector whose components are \gls{iid} by Gaussian random variables with zero mean and unit variance: $\mathbf{Z}\sim \mathcal{N}(0,\mathbf{I}_{n})$ independent of $\mathbf{X}$, where $\mathbf{I}_n$ denotes the $n \times n$ identity matrix. 

Furthermore, we define a list $\mathcal{C} = \left\lbrace  1,\cdots ,M \right\rbrace $ of $M$ equiprobable messages to be transmitted. Thus, an $(n,M,\epsilon)$-code has an encoding function $f: \mathcal{C}  \longmapsto \mathcal{A}^{n}$ in which a message $m\in \mathcal{C}$ is chosen and returns a codeword $\mathbf{X}^{m} = \left[X_{1}^{m}, \cdots, X_{n}^{m}\right]$. All of these input sequences satisfy a maximal-power constraint such that:
\begin{equation}
\Vert \mathbf{X}^{m} \Vert^{2} \leq n PT,~~ m = 1,\cdots,M\,,
\label{eq:0.0.1}
\end{equation}
where $P$ is the average received power and $T$ denotes the symbol duration. In addition, consider a decoder $g: \mathcal{B}^{n} \longmapsto\mathcal{C}$ whose average probability of error does not exceed the \gls{bler} $\epsilon$:
\begin{equation}
{1 \over M}\sum\limits_{m = 1}^M {{P_{\mathbf{Y}|m}}\left( {g\left( \mathbf{Y} \right) \ne m|M = m} \right) \le \epsilon }
\end{equation}
where $\text{P}_{\mathbf{Y}\mid m}$ is the conditional probability that the decoder $g(\mathbf{Y})$ picks up the wrong message when the actual $m$ was transmitted.

The maximum achievable code rate $R^{*}(\epsilon,n) = \log_{2}(M)/n$ (bits per channel use) satisfying a required \gls{bler} $\epsilon$ given a \gls{snr} budget $ E_{s}/N_{0} = PT$ can be determined for finite-blocklength coding in \gls{awgn} channels. It is given by the following theorem \cite{Polyanskiy2010}.
 
\begin{theorem}[Polyanskiy \textit{et al}. 2010]
\label{th:2}
For a discrete-time \gls{awgn} channel with a \gls{snr} equal to $PT$, there exists a $(n,M,\epsilon)$-code such that the maximum achievable code rate $R^{*}(\epsilon,n)$ for equal-power and maximal-power constraints is given by:
\begin{equation}
  R^{*}(\epsilon,n) \leq C-\sqrt{\frac{V}{n}}Q^{-1}(\epsilon)+\frac{1}{2n}\log_{2}(n)+O(1)\,,
\end{equation}
where $Q^{-1}(\cdot)$ is the inverse complementary Gaussian CDF function and $C$ is Shannon's capacity formula in bits per channel use:
\begin{equation}
C = \frac{1}{2}\log_{2}(1+PT)
\end{equation}
assuming an \gls{awgn} channel with zero mean and unit variance. $ V$ is the channel dispersion written as:
\begin{equation}
V = \frac{PT}{2} \frac{PT+2}{(PT+1)^{2}} \log_{2}^{2}e\,.
\end{equation}
\end{theorem}

The channel dispersion quantifies, under equal capacity, the stochastic variability of the channel relative to a deterministic channel. By assuming that the input sequence is encoded such that $\mathbf{X}^{m} \in \mathbb{R}^{n}$ for $m = 1,\cdots, M$ and the received signal is $\mathbf{Y} \in \mathbb{R}^{n}$, we assume that each component of the vector $\mathbf{X}^{m}$ is transmitted in parallel using an \gls{ofdm}-like signal. 
 
\subsection{Transmission using OFDM-like Signals}
Consider a message $m$ chosen for transmission. Each component of the codeword $\mathbf{X}^{m}= \{ \mathop X\nolimits_1^m ,...,\mathop X\nolimits_n^m \} $ is simultaneously transmitted in parallel over the channel via an \gls{ofdm}-like signal such that:
\begin{equation}
s_{m}(t) = \sum\limits_{i = 1}^{n} X^{m}_{i}\varphi_{i}(t), \hspace{0.5cm} 0 \leq t \leq T\,,
\end{equation}
where $\varphi_{i}(t)$ is an orthonormal basis spanning the vector-space of signals. We consider that the decoder receives an output sequence $\mathbf{Y}_{\tau} = \left\lbrace Y_{1,\tau},\cdots, Y_{n,\tau} \right\rbrace $ for all $\tau \in \left[ 0,T \right]$ such that:
\begin{equation}
Y_{i,\tau} = \int_{0}^{\tau}(s_{m}(t)+z(t))\varphi_{i}^{*}(t) dt\,,
\end{equation}
where $z(t)$ is a zero-mean white Gaussian process with power spectral density $N_0/2$. This condition allows the design of an optimal decision rule minimizing the time needed ($\tau \leq T$) to make a correct decision while maintaining a \gls{bler} that does not exceed $\epsilon$.

In \gls{awgn} channels, we assume that $\forall~\delta\tau>0$ the differential $Y_{i,\tau+\delta \tau}-Y_{i,\tau}$ is \gls{iid} $\forall~i = 1, \cdots, n$, and consider the case of equiprobable signals, i.e., $P_m = 1/M$ for all $m$. It follows that the decoder can do early detection based on a sequential test by considering a threshold $ S_{m} \geq 0$:
\small
\begin{equation}
g(\mathbf{Y}_{\tau})  =
  \begin{cases}
   m        & \text{if } \exists~\mathbf{X}^{m}~\text{s.t.}~P(m|{\mathbf{Y}_{\tau} }) > {S_m}\,, \\
   \text{wait for}~\mathbf{Y}_{\tau+\delta\tau}        & \text{otherwise}\,.
  \end{cases}
\label{eq:3.00.00}
\end{equation}
\normalsize
The decoder makes a decision as soon as the posterior probability $P(m|{\mathbf{Y}_{\tau} })$ reaches a threshold $S_{m}$. If the threshold is not reached before the end of the transmitted symbol, the decision is made at $t=T$.

According to the guidelines set by the binary sequential probability ratio test (SPRT) \cite{Au2016}, the threshold $S_m$ can be defined as follows:
\small
\begin{equation}
  {S_m} = {1 \over {1 + M\sum\limits_{\scriptstyle m' = 1 \hfill \atop 
        \scriptstyle m' \ne m \hfill} ^M {Q\left( {\sqrt {{{\mathop d\nolimits_{mm'}^2 } \over {2{N_0}}}} } \right)} }}\,,
\end{equation}
\normalsize
where $Q(\cdot)$ is the complementary CDF of a Gaussian random variable, and ${\mathop d\nolimits_{mm'}^2 }$ is the squared distance between two different codewords.

To reduce the probability of false decision, the cross product between two arbitrary \gls{ofdm}-like signals $s_{m}(t)$ and $s_{m^{'}}(t)$ with $m^{'} \in \mathcal{C}$ and $m \neq m^{'}$ is assumed to be approximately equal to zero. In other words $\forall \tau \in \left[  0,T  \right] $, the \gls{ofdm}-like signals must satisfy:
\small
\begin{align*}
d_{mm^{'}}^{2}(\tau) &= \int_{0}^{\tau} \vert s_{m}(t)-s_{m^{'}}(t) \vert^{2} dt\\ 
&\approx \Vert s_{m}(\tau) \Vert^{2} +\Vert s_{m^{'}}(\tau) \Vert^{2} - \underbrace{2\mathcal{R}\left\lbrace\int_{0}^{\tau}  s_{m}(t)s_{m^{'}}^{*}(t) dt\right\rbrace }_{\approx 0}\,,
\end{align*}
\normalsize
where, subject to a maximal-power constraint, each $s_m(\tau)$ satisfies (cf. Eq.\,\eqref{eq:0.0.1}):
\begin{equation}
  \label{eq:1.0.0.0} 
  \begin{split}
    \Vert s_{m}(\tau) \Vert^{2} &= \int_{0}^{\tau} s_{m}(t)s_{m}^{*}(t) dt \leq n P T \\
    &\text{for all }m = 1,\cdots, M\,,
  \end{split}
\end{equation}
where $\mathcal{R}\left\lbrace \cdot \right\rbrace$ denotes the real part of a complex number and $(\cdot)^{*}$ is the complex conjugate. A decoder able to make correct decisions before the end of the transmitted symbol reduces the signaling delay. Furthermore, \eqref{eq:1.0.0.0} shows that the decoder aims to detect the message using less signal energy than stated in the right-hand-side bound of \eqref{eq:0.0.1}. 

The orthogonality of an \gls{ofdm} transmission depends on the space between subcarrier. This space must be equal to $\nicefrac{1}{T}$. As mentioned above, the proposed early detection scheme makes a sequential decision before the end of the symbol duration $T$. As a consequence, the orthogonality of the signals is not preserved, and the distance between two \gls{ofdm} signals becomes nonlinear. Therefore, to reduce latency and preserve the orthogonality through the proposed early-detection scheme, ${\mathop d\nolimits_{mm'}^2 }$ must be linear for all $m \ne {m'}$ in an arbitrary codebook. An efficient solution to linearize these distances is to employ a precoding random matrix, in particular, a Hadamard transform, which renders the early-detection scheme feasible over \gls{ofdm} \cite{Au2016}.

\section{Optimal Latency in One-Hop Communication Over AWGN Channels}
\label{sect:III}
Consider a channel code with a blocklength of $n$ channel uses, and with a symbol duration $T$. A  decoder that makes its decision at the end of a transmitted symbol has latency: 
\begin{equation}
  L = nT\,.
  \label{eq:3.0.0}
\end{equation}
In this work, we consider a transmission with a fixed duration, but where the receiver operates at $\tau \in [0,T]$. Therefore, we are concerned by the minimal average $\mathbb{E}\left[ \tau \right]$ needed to decode one of the $M$ possible messages for a targeted \gls{bler} $\epsilon$.

\subsection{On Minimal Latency using an Optimal Stopping-Time Rule}
The optimal average latency can be achieved through a genie-aided detector, i.e.,  a perfect error-detecting code is assumed \cite{Au2016}. Consider a sequential test characterized by:
\begin{subequations}
  \begin{align}
    \tau = \inf\left\lbrace t: P(m|{\mathbf{Y}_{\tau} }) > {S_m} \right\rbrace\,, \\
    \delta = \hat{m},~  \text{where}~ \hat{m} = \argmax\limits_{1\leq m\leq M}P(m|{\mathbf{Y}_{\tau} })\,,
  \end{align}
  \label{eq:1.1} 
\end{subequations} 
where the stopping time $\tau$ is the minimal time in which the decoder has made a decision when $s_{m}(t)$ was transmitted, and $\delta$ is the final decision rule. The decoder makes its decisions as soon as the posterior probability reaches a predetermined threshold $S_{m}$. The probability of having a correct decision given a message $m$ is denoted by $p({\tau }|{s_m}(t))$. Since $\tau$ is a random variable, the average latency is: 
\begin{equation}
  \mathbb{E}\left[ \tau \right]= \frac{1}{M} \sum\limits^{M}_{m = 1} \mathbb{E}\left[ {{\tau}|{s_m}(t)} \right]\,,
\end{equation}
where $\mathbb{E}\left[\cdot \right]$ is the expectation value. Assuming the message has a uniform distribution, the conditional average of $\tau$ given $M=m$ is provided by:
\begin{equation}
  \mathbb{E}\left[ {{\tau}|{s_m}(t)} \right] = \int_{0}^{T}\tau p({\tau}|{s_m}(t))  d\tau\,.
\end{equation}
Similarly, the probability of having a correct decision at a given $\tau$ for an $(n,M,\epsilon)$-code is given by:
\begin{equation}
  p(\tau )  = \frac{1}{M} \sum\limits^{M}_{m = 1} p({\tau }|{s_m}(t))\,.
\end{equation}
In an optimal early-detection scheme the average \gls{bler} must be:
\begin{equation}
  \epsilon = 1- \int_{0}^{T} p(\tau )  d\tau\,.
\end{equation}

\subsection{Minimal Latency of Channel Codes in the Finite Block-Length Regime over AWGN Channels}
\label{sect:III:MinLatencyFinite}
For synchronous detection, the minimal latency can be obtained for an $(n,M,\epsilon)$-code by taking the minimum blocklength $n$ needed to achieve a required $\epsilon$ given $k=\log_2 M$ information bits and a code rate $R$. Using the maximum achievable code rate given by Theorem~\ref{th:2}, the minimum blocklength $n(\epsilon,\log_{2} M)$ can be approximated by:
\begin{equation}
  n(\epsilon,\log_{2} M) \approx   \left( \frac{Q^{-1}(\epsilon)}{C-R^{*}(\epsilon,n)}\right)^{2} V,
\end{equation}
and the latency given by $L_{\text{SD}} = n(\epsilon,\log_{2} M)T$. 

In order to derive the minimal latency of an optimal early-detection scheme, let $\epsilon(\tau)$ be the \gls{bler} as a function of $\tau$. From Theorem~\ref{th:2}, it can be expressed as:
\begin{equation}
  \epsilon(\tau) \approx Q\left( \frac{C(\tau)-R+ \frac{1}{2n}\log_{2} n}{\sqrt{V(\tau)/n}}\right)\,,
\end{equation}
where $C(\tau)$ and $V(\tau)$ are functions of $P\tau$. By letting $\epsilon(\tau-\delta \tau)$ and $\epsilon(\tau)$ be the average \gls{bler} when decisions are made at $\tau-\delta \tau$ $\forall \delta \tau > 0$ and $\tau$ respectively, where $\epsilon(\tau-\delta \tau) \geq \epsilon(\tau)$, the probability of having a correct decision at $\tau$ is:
\begin{align*}
    p(\tau ) &= (1-\epsilon(\tau))-(1-\epsilon(\tau-\delta \tau))\\
    &=  \epsilon(\tau-\delta \tau)-\epsilon(\tau)\,.
\end{align*}
By letting $\delta \tau \rightarrow 0$, the distribution of $\tau$ is equal to the differential of the \gls{bler}: 
\begin{equation}
    p(\tau ) = - d\epsilon(\tau)\,.
  \label{eq:3.1.1}
\end{equation}
The optimal average latency for this early-detection scheme can be computed by taking the expectation value for a given blocklength $n$ and code rate $R$ using \eqref{eq:3.1.1}. Thus, the optimal average latency in an early-detection scheme is $L_{\text{ED}} = n\mathbb{E}\left[\tau\right]$.

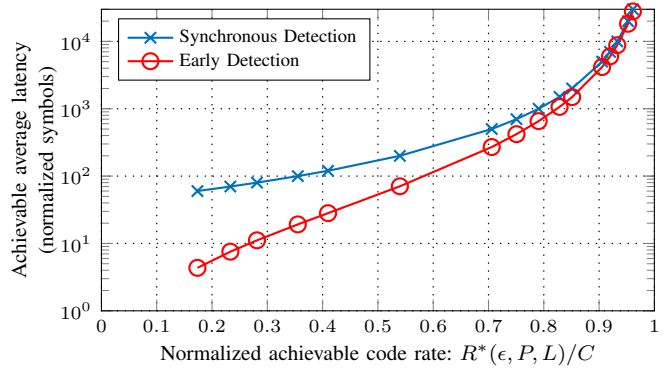
\begin{figure}[t]
  \centering
  \input{figures/SD_vs_ED_onehop}\vspace*{-1.2em}
  \caption{Optimal achievable latency in one-hop communications. The \gls{bler} $\epsilon = 10^{-12}$, and the \gls{snr} link budget is of $5$\,dB.}
  \label{fig:6.1.0}
\end{figure}

Fig.\,\ref{fig:6.1.0} shows results on the minimal latency in one-hop communications using either synchronous detection or an optimal early detection at any fixed normalized achievable code rate. These results are for a targeted \gls{bler} $\epsilon = 10^{-12}$ and an \gls{snr} link budget of $5$\,dB, and latency is expressed in terms of normalized symbols ($L/T$). The proposed early-detection scheme is shown to have lower latency than synchronous detection as the decoder makes a reliable decision before the end of the transmitted symbols. The early-detection scheme is also shown to loose its advantage over synchronous detection as the channel codes operate near capacity. Therefore, such a scheme is more efficient for short messages. 

In the following, we show that our findings can be applied to typical multi-hop scenarios where short messages are used.

\section{Optimal Latency for Low-Traffic Multi-hop Systems}
\label{sect:IV}
In time-sensitive applications, radio devices require time-synchronized low-latency services \cite{Schaich2014}. We apply the early-detection strategy for low-traffic multi-hop systems by retransmitting short messages as soon as they are correctly decoded. Consider $h$ hops that employ a \gls{df} relaying scheme with synchronous detection where messages with latency $L$ (as provided by \eqref{eq:3.0.0}) are transmitted. Latency is defined as:
\begin{equation}
  L_{\text{SD-DF}} = Lh\,.
  \label{eq:4.0.0bis}
\end{equation}   
If a relay makes correct decisions before the end of the transmitted symbol using an early-detection scheme and retransmits the message as soon as possible, latency is reduced.

\begin{theorem}
  \label{th:3}
  In low-traffic multi-hop systems, the minimal latency of \gls{df} relaying schemes using early-detections is: 
  \begin{equation}  
   \mathbb{E}\left[ L_{\text{ED-DF}}\right] \leq L_{\text{SD-DF}}\,.
  \end{equation}
\end{theorem}

\textit{Proof:} Assuming that the $h$ \gls{df} relays employ an $(n,M,\epsilon)$-code along with an optimal early-detection scheme,
if a source transmits an arbitrary message to a relay $R_{\ell}$ with a decoder that satisfies \eqref{eq:3.00.00} at $\tau$ as in \eqref{eq:1.1}, the hop will retransmit the message through an \gls{ofdm}-like signal to the next hop with lower latency than a scheme with synchronous detection.

In order to calculate the overall latency for $h$ hops, we consider a source $S$ that transmits an arbitrary message encoded by an $(n,M,\epsilon)$-code, where latency $L_{0} = nT$. Next, a relay $R_{1}$ performs an optimal early-detection scheme, by which the decision has been made at $\tau_{1} \leq T$. Therefore, the signaling delay is $L_{1} = L_{0}+\tau_{1}n$, and so on. Assuming an \gls{iid} random sequence $\tau_{1}, \cdots, \tau_{h} \leq T$ that satisfies \eqref{eq:1.1}, the overall average latency for $h$ hops is:
  \begin{align*}
    L_{\text{ED-DF}} &= {L_0} + {\tau _1}n + {\tau _2}n + ... + {\tau _{h - 1}}n\,,\\
    \mathbb{E}\left[ L_{\text{ED-DF}} \right] &= nT + \mathbb{E}\left[\sum\limits_{i = 1}^{h-1} \tau_{i}n \right]\\
                                          &= nT + (h-1)n\mathbb{E}\left[\tau\right]\leq hnT\\
                                          &\leq L_{\text{SD-DF}}\,.
  \end{align*}
The above proves that early-detection schemes can reduce latency in multi-hop systems using \gls{df} relaying schemes. \qed

\section{Optimal Latency for Phased Continuous-Transmission Multi-hop Links}
\label{sect:V}

\begin{figure}[t]
  \centering
  \includegraphics[width=\columnwidth]{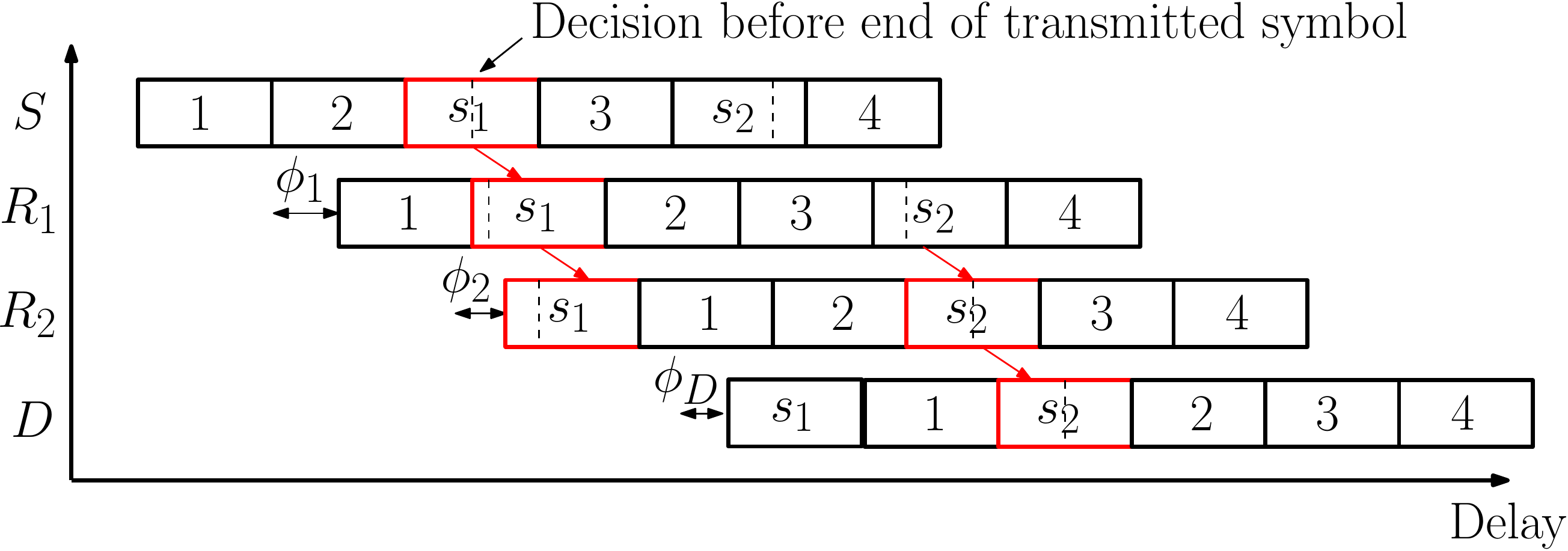}\vspace*{-1.0em}
  \caption{Phased continuous-transmission multi-hop links using early-detection schemes. $1$, $2, \cdots$ denotes the queued traffic, $\phi_i$ additional delays due to non-synchronous multi-hop links, and $s_{\text{x}}$ short message blocks. Blocks in red designate retransmissions ahead of queued traffic.}
  \label{fig:5.1.2}
\end{figure}

Vehicle-to-vehicle or machine-to-machine multi-hop links are generally non synchronous, resulting in greater latency. However, assuming continuous transmission with variable-length blocks, a relay that would retransmit correct short messages ahead of queued traffic would reduce latency. Fig.\,\ref{fig:5.1.2} illustrates the proposed strategy. Assuming a random phase $\phi_{i}$ at each hop due to non-synchronous links between hops, the total average latency for \gls{df}-relaying schemes with synchronous detection is:
\begin{equation}
  \begin{split}
    L_{\text{CTSD}} &= \mathbb{E}\left[\sum\limits_{i = 1}^{h} \phi_{i}nT+ \sum\limits_{i = 1}^{h} nT\right]\\
    &= \left( \mathbb{E}\left[\phi \right]+1\right)  L_{\text{SD-DF}}\,.\\
  \end{split}
  \label{eq:5.1.1}
\end{equation}  

For \gls{df}-relaying using an optimal early-detection scheme, the overall average latency is given by the following theorem. 
\begin{theorem}
  \label{th:4}
  For phased continuous-transmission multi-hop links using a \gls{df}-relaying scheme with a random phase $\phi_{i}$ at each hop, the optimal average latency is bounded by:
  \begin{equation}
    \mathbb{E}\left[\phi \right] L_{\text{SD-DF}} \leq L_{\text{CTED}} \leq \mathbb{E}\left[\phi \right] L_{\text{SD-DF}}+\mathbb{E}\left[ L_{\text{ED-DF}} \right]\,.
    \label{eq:5.1.2}
  \end{equation}
\end{theorem}

\textit{Proof:} Assuming that the random phase $\phi$ and the time needed to make a correct decision $\tau$ are independent, this theorem is the immediate consequence of Theorem~\ref{th:3} and of \eqref{eq:5.1.1}. The lower bound in \eqref{eq:5.1.2} implies that all the relays made their decision before the phase $\phi$. The minimal average latency $L_{\text{CTED}} $ is equal to its upper bound if all relays made their decisions after the phase. The upper bound in \eqref{eq:5.1.2} is itself upper bounded by \eqref{eq:5.1.1}. \qed

Therefore, in phased continuous-transmission multi-hop links, the use of early detection can reduce latency.

\section{Results and Discussion}
\label{sect:VI}
In this section, we present results for both low-traffic multi-hop systems and for phased continuous-transmission multi-hop links. The proposed optimal early-detection scheme is compared to synchronous detection for various $(n,M,\epsilon)$-codes. We compute the exact optimal latency for low-traffic multi-hop systems with relays that employ an optimal early-detection scheme via Theorem~\ref{th:3}, and we do the same for phased continuous-transmission multi-hop links via Theorem~\ref{th:4}. As in Section~\ref{sect:III:MinLatencyFinite}, the average latency is expressed in terms of normalized symbols.

\begin{figure}[t]
  \centering
  \input{figures/res_multi_hop1}\vspace*{-1.2em}
  \caption{Optimal achievable average latency of phased continuous-transmission multi-hop links and of low-traffic multi-hop systems using either synchronous detection (SD) or early detection (ED). The number of hops $h = 10$, the \gls{bler} $\epsilon = 10^{-12}$, and the \gls{snr} link budget is of 5\,dB.}
  \label{fig:6.1.1}
\end{figure}
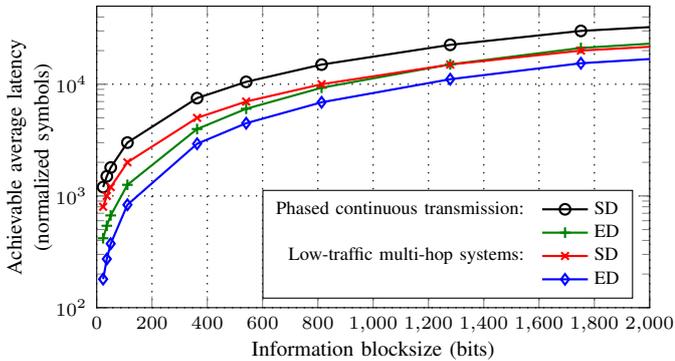

Fig.\,\ref{fig:6.1.1} shows the optimal achievable average latency for various information blocksizes of both phased continuous-transmission multi-hop links and low-traffic multi-hop systems. Results are for $10$ hops, an \gls{snr} link budget per hop of $5$\,dB, a \gls{bler} $\epsilon = 10^{-12}$, and the random phase $\phi$ is assumed to be uniformly distributed over $\left[0, 1\right]$, thus $\mathbb{E}\left[\phi \right]= 1/2$.
As expected, the latency is shown to increase with the information blocksize.
More importantly, compared to synchronous detection, the proposed early-detection scheme shows an improvement in latency in both multi-hop scenarios.

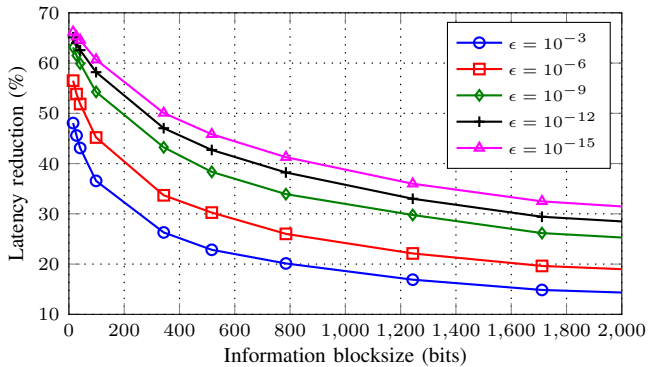
\begin{figure}[t]
  \centering
  \input{figures/res_multihop_epsil2}\vspace*{-1.2em}
  \caption{Latency reduction for phased continuous-transmission multi-hop links using an optimal early-detection scheme with various \glspl{bler} $\epsilon$. The number of hops $h = 4$, and the \gls{snr} link budget is of 5\,dB.}
  \label{fig:6.1.2}
\end{figure}

The proposed early-detection scheme can significantly reduce latency, especially for short codes with a very-small \gls{bler} ($\epsilon \rightarrow 0$).
Fig.\,\ref{fig:6.1.2} shows the latency reduction for various \glspl{bler} $\epsilon$ and information blocksizes with a fixed number of hops $h = 4$ and an \gls{snr} link budget of $5$\,dB. The latency reduction is calculated as $100 \times \left(1-\nicefrac{L_{\text{CTED}}}{L_{\text{CTSD}}}\right)$.
Compared to synchronous detection, it can be seen that the proposed early-detection scheme offers a latency reduction in the range of 20\% to 65\%. For any fixed \gls{bler} $\epsilon$, codes with a smaller information blocksize have the most significant latency reduction. For a fixed information blocksize, a smaller \gls{bler} $\epsilon$ leads to greater latency reduction. 

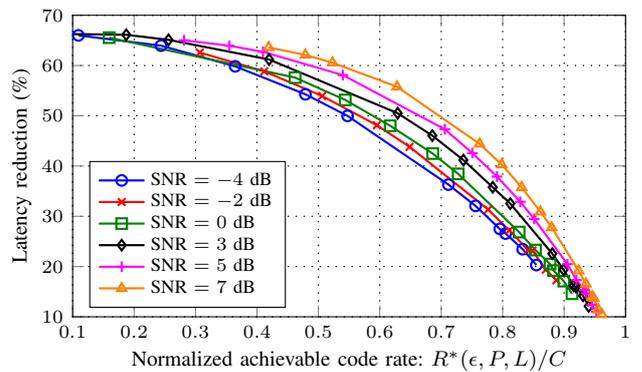
\begin{figure}[t]
  \centering
  \input{figures/res_multihop_SNR}\vspace*{-1.2em}
  \caption{Latency reduction for normalized achievable code rates with various \gls{snr} link budgets. The number of hops $h = 4$, and the \gls{bler} $\epsilon = 10^{-12}$.}
  \label{fig:6.1.3}
\end{figure}

Fig.\,\ref{fig:6.1.3} shows the latency reduction for various normalized achievable code rates and  \gls{snr} link budgets, where the number of hops $h = 4$ and the \gls{bler} $\epsilon = 10^{-12}$. It can be seen that the proposed early-detection scheme is particularly suitable for short blocklengths as the latency reduction diminishes as the code rate tends to capacity. Lower-rate codes benefit the most, with a latency reduction that culminates at little over 65\%. 
Fig.\,\ref{fig:6.1.3} also shows a latency reduction well above 40\% over a very-large range of code rates, especially if the \gls{snr} budget link per hop can be increased.

\section{Conclusion}
\label{sect:VII}
In this work, we described a strategy for one-hop and linear multi-hop systems with short messages that provably reduces latency without compromising reliability.
The key element is an early-detection scheme based on a sequential-test technique where short messages are detected and retransmitted before being completely received, and where the test threshold ensures a targeted error probability.
Compared to decode-and-forward relays that use synchronous detection, we have proved that the proposed scheme has the lowest overall latency for short messages.
In addition, our results showed that over continuous-transmission links, the overall latency of the proposed scheme can be reduced by up to 66\% while maintaining a block-error rate as low as $10^{-15}$.
The proposed early-detection scheme offers promising results over the AWGN channel, where it was shown to be suitable for ultra-reliable and extremely low-latency communications.
Lastly, extending the proposed scheme to fading channels would be of interest for future work.

\balance
\bibliographystyle{IEEEtran}
\bibliography{IEEEabrv,ConfAbrv,Data_base_latency}

\end{document}

%% file: figures/SD_vs_ED_onehop.tex
\definecolor{mycolor1}{rgb}{0.00000,0.44700,0.74100}%
\begin{tikzpicture}

  \pgfplotsset{
    grid style = {
      dash pattern = on 0.05mm off 1mm,
      line cap = round,
      black,
      line width = 0.5pt
    },
    title style = {font=\fontsize{7pt}{7.2}\selectfont},
    label style = {font=\fontsize{8pt}{7.2}\selectfont},
    tick label style = {font=\fontsize{7pt}{7.2}\selectfont}
  }

  \begin{semilogyaxis}[%
    width=\columnwidth,
    height=\plotfigureheight\columnwidth,
    grid=major,
    xmin=0, xmax=1,
    xtick={0,0.1,...,1.1},
    xlabel={Normalized achievable code rate: $R^{*}(\epsilon,P,L)/C$},
    xlabel style={yshift=0.6em},%
    ymin=1, ymax=30000,
    ylabel={\shortstack{Achievable average latency\\(normalized symbols)}},
    ylabel style={yshift=-1.0em},
    legend pos={north west},
    legend style={
      anchor={north west},
      cells={anchor=west},
      column sep=0mm,
      row sep=-0.5mm,
      font=\fontsize{7pt}{7.2}\selectfont,
      mark size=3.0pt,
      mark options=solid
    },
    mark size=3.0pt,
    mark options=solid,
    every axis plot/.append style={thick},
    ]
    \addplot [color=mycolor1,mark=x]
    table[row sep=crcr]{%
      0.173627446966806	60\\
      0.233185597584884	70\\
      0.281310045623787	80\\
      0.35512194933836	100\\
      0.40980648808702	120\\
      0.539744560715088	200\\
      0.70587569997265	500\\
      0.750616042004896	700\\
      0.790703676880378	1000\\
      0.828574051202422	1500\\
      0.851244966980092	2000\\
      0.905428069695878	5000\\
      0.919949514028598	7000\\
      0.932928762365751	10000\\
      0.952464014279562	20000\\
      0.961144481959429	30000\\
    };
    \addlegendentry{Synchronous Detection}

    \addplot [color=red,mark=o]
    table[row sep=crcr]{%
      0.173627446966806	4.34485319720607\\
      0.233185597584884	7.5545606993237\\
      0.281310045623787	11.1322194324471\\
      0.35512194933836	19.1843815574852\\
      0.40980648808702	28.1913965372281\\
      0.539744560715088	70.5567330543714\\
      0.70587569997265	270.170650506489\\
      0.750616042004896	419.275118807044\\
      0.790703676880378	654.706816515626\\
      0.828574051202422	1065.42404114236\\
      0.851244966980092	1489.91107479934\\
      0.905428069695878	4161.15302925747\\
      0.919949514028598	5996.92143085124\\
      0.932928762365751	8789.70721974716\\
      0.952464014279562	18263.8519603288\\
      0.961144481959429	27860.1730377921\\
    };
    \addlegendentry{Early Detection}

  \end{semilogyaxis}

\end{tikzpicture}%

%% file: figures/res_multi_hop1.tex
\definecolor{darkgreen}{RGB}{0, 128, 0}
\begin{tikzpicture}

  \pgfplotsset{
    grid style = {
      dash pattern = on 0.05mm off 1mm,
      line cap = round,
      black,
      line width = 0.5pt
    },
    title style = {font=\fontsize{7pt}{7.2}\selectfont},
    label style = {font=\fontsize{8pt}{7.2}\selectfont},
    tick label style = {font=\fontsize{7pt}{7.2}\selectfont}
  }

  \begin{semilogyaxis}[%
    width=\columnwidth,
    height=\plotfigureheight\columnwidth,
    grid=major,
    xmin=0, xmax=2000,
    xtick={0,200,...,10000},
    scaled x ticks=false,
    xlabel={Information blocksize (bits)},
    xlabel style={yshift=0.6em},%
    ymin=1e2, ymax=5e4,
    ylabel={\shortstack{Achievable average latency\\(normalized symbols)}},
    ylabel style={yshift=-1.0em},
    legend pos={south east},
    legend columns=2,
    legend style={
      anchor={south east},
      cells={anchor=west},
      column sep=0mm,
      row sep=-0.5mm,
      font=\fontsize{7pt}{7.2}\selectfont,
      mark size=2.0pt,
      mark options=solid
    },
    mark size=2.0pt,
    mark options=solid,
    every axis plot/.append style={thick},
    ]
    
    \addlegendimage{empty legend}
    \addlegendentry[anchor=east]{Phased continuous transmission:}
    \addplot [color=black, mark=o]
    table[row sep=crcr]{%
      23.1503900471334	1200.73915940986\\
      36.5309192178481	1501.05901102864\\
      50.5874933582085	1797.1577920296\\
      111.045599870647	3003.56607453504\\
      363.062438432575	7515.32007478451\\
      540.504067169971	10503.0708435907\\
      813.386280380287	14987.4813660509\\
      1278.51454071799	22509.7015569306\\
      1751.32858902622	29999.9851469292\\
      4657.00863228216	74969.4318174976\\
      6624.37819201648	104981.279452395\\
      9596.91320614996	149885.459733134\\
      10591.6684799872	164850.212439624\\
      14584.0821747206	225057.677739376\\
      19595.7394514085	300234.4151675\\
    };
    \addlegendentry{SD}
    \addlegendimage{empty legend}
    \addlegendentry{}
    \addplot [color=darkgreen, mark=+]
    table[row sep=crcr]{%
      23.1503900471334	418.065079469059\\
      36.5309192178481	541.012757566112\\
      50.5874933582085	669.439322044901\\
      111.045599870647	1257.92926716698\\
      363.062438432575	3972.29464703024\\
      540.504067169971	6028.76955857789\\
      813.386280380287	9298.20094824676\\
      1278.51454071799	15056.5260251753\\
      1751.32858902622	21104.024837326\\
      4657.00863228216	59606.6696148756\\
      6624.37819201648	86326.8172327727\\
      9596.91320614996	127337.448203019\\
      10591.6684799872	140813.406331563\\
      14584.0821747206	196863.730974908\\
      19595.7394514085	266836.14708482\\
    };
    \addlegendentry{ED}

    \addlegendimage{empty legend}
    \addlegendentry[anchor=east]{Low-traffic multi-hop systems:}
    \addplot [color=red, mark=x]
    table[row sep=crcr]{%
      23.1503900471334	800\\
      36.5309192178481	1000\\
      50.5874933582085	1200\\
      111.045599870647	2000\\
      363.062438432575	5000\\
      540.504067169971	7000\\
      813.386280380287	10000\\
      1278.51454071799	15000\\
      1751.32858902622	20000\\
      4657.00863228216	50000\\
      6624.37819201648	70000\\
      9596.91320614996	100000\\
      10591.6684799872	110000\\
      14584.0821747206	150000\\
      19595.7394514085	200000\\
    };
    \addlegendentry{SD}
    \addlegendimage{empty legend}
    \addlegendentry{}

    \addplot [color=blue, mark=diamond]
    table[row sep=crcr]{%
      23.1503900471334	180.159058790588\\
      36.5309192178481	272.811982119821\\
      50.5874933582085	373.705469054691\\
      111.045599870647	833.830254302544\\
      363.062438432575	2931.19032190322\\
      540.504067169971	4476.02629476296\\
      813.386280380287	6891.2068470685\\
      1278.51454071799	11088.208397084\\
      1751.32858902622	15404.6187361874\\
      4657.00863228216	42444.4166441664\\
      6624.37819201648	60993.101596016\\
      9596.91320614996	89110.917809178\\
      10591.6684799872	98543.5207352076\\
      14584.0821747206	136567.180721807\\
      19595.7394514085	184362.049820498\\
    };
    \addlegendentry{ED}

  \end{semilogyaxis}

\end{tikzpicture}%

%% file: figures/res_multihop_epsil2.tex
\definecolor{darkgreen}{RGB}{0, 128, 0}
\definecolor{mycolor1}{rgb}{1.00000,0.00000,1.00000}%

\begin{tikzpicture}

  \pgfplotsset{
    grid style = {
      dash pattern = on 0.05mm off 1mm,
      line cap = round,
      black,
      line width = 0.5pt
    },
    title style = {font=\fontsize{7pt}{7.2}\selectfont},
    label style = {font=\fontsize{8pt}{7.2}\selectfont},
    tick label style = {font=\fontsize{7pt}{7.2}\selectfont},
  }

  \begin{axis}[%
    width=\columnwidth,
    height=\plotfigureheight\columnwidth,
    grid=major,
    xmin=0, xmax=2000,
    xtick={0,200,...,10000},
    scaled x ticks=false,
    xlabel={Information blocksize (bits)},
    xlabel style={yshift=0.6em},%
    ymin=10, ymax=70,
    ytick={0,10,...,80},
    ylabel={Latency reduction (\%)},
    ylabel style={yshift=-1.6em},
    legend style={
      at={(.98,.97)},
      anchor={north east},
      cells={anchor=west},
      column sep=0mm,
      row sep=-0.5mm,
      font=\fontsize{6pt}{7.2}\selectfont,
      mark size=2.0pt,
      mark options=solid
    },
    mark size=2.0pt,
    mark options=solid,
    every axis plot/.append style={thick},
    ]

    \addplot [color=blue, mark=o]
    table[row sep=crcr, y expr=100*(1-\thisrowno{1})]{%
      15.1181786195608	0.519368941132182\\
      27.5506338369966	0.544109834053093\\
      40.750083606356	0.568956531477686\\
      98.3455584910656	0.634441976278131\\
      342.981909863643	0.737028242316752\\
      516.744465349849	0.771587570936652\\
      784.988124538484	0.79875321694273\\
      1243.73404499377	0.831013635256667\\
      1711.16753188835	0.851466456115011\\
      4593.50842538426	0.902766572414459\\
      6549.24373396621	0.919291715506439\\
      9507.11035234144	0.930570354309315\\
      10497.4824523219	0.93256809405093\\
      14474.0965900823	0.943144705424291\\
      19468.7390376127	0.947479353683361\\
    };
    \addlegendentry{$\epsilon = 10^{-3}$}

    \addplot [color=red, mark=square]
    table[row sep=crcr, y expr=100*(1-\thisrowno{1})]{%
      15.1181786195608	0.434973320279241\\
      27.5506338369966	0.46159620619778\\
      40.750083606356	0.481439153313043\\
      98.3455584910656	0.548088033623444\\
      342.981909863643	0.663122953452509\\
      516.744465349849	0.697494121938918\\
      784.988124538484	0.739852515414614\\
      1243.73404499377	0.778839438776206\\
      1711.16753188835	0.803656995986297\\
      4593.50842538426	0.868007084662976\\
      6549.24373396621	0.884933022579687\\
      9507.11035234144	0.904420030298368\\
      10497.4824523219	0.906913846211371\\
      14474.0965900823	0.91798801976338\\
      19468.7390376127	0.933366439431832\\
    };
    \addlegendentry{$\epsilon = 10^{-6}$}

    \addplot [color=darkgreen, mark=diamond]
    table[row sep=crcr, y expr=100*(1-\thisrowno{1})]{%
      15.1181786195608	0.368640265419053\\
      27.5506338369966	0.384996788140346\\
      40.750083606356	0.40113210403454\\
      98.3455584910656	0.457400123128509\\
      342.981909863643	0.567626796099846\\
      516.744465349849	0.616576284068543\\
      784.988124538484	0.660792431555827\\
      1243.73404499377	0.702313730127961\\
      1711.16753188835	0.738344566704538\\
      4593.50842538426	0.825679215196984\\
      6549.24373396621	0.851495684977704\\
      9507.11035234144	0.868208078715044\\
      10497.4824523219	0.872453443553484\\
      14474.0965900823	0.891284482262219\\
      19468.7390376127	0.90436207887257\\
    };
    \addlegendentry{$\epsilon = 10^{-9}$}

    \addplot [color=black, mark=+]
    table[row sep=crcr, y expr=100*(1-\thisrowno{1})]{%
      15.1181786195608	0.348886948650519\\
      27.5506338369966	0.359855718149721\\
      40.750083606356	0.374265640315248\\
      98.3455584910656	0.418121433343233\\
      342.981909863643	0.529589094456033\\
      516.744465349849	0.572905158782815\\
      784.988124538484	0.617732589333994\\
      1243.73404499377	0.66999299551332\\
      1711.16753188835	0.705847491667448\\
      4593.50842538426	0.79831193697475\\
      6549.24373396621	0.823982251044603\\
      9507.11035234144	0.845350925556353\\
      10497.4824523219	0.856414769618695\\
      14474.0965900823	0.874761824724311\\
      19468.7390376127	0.888875982655863\\
    };
    \addlegendentry{$\epsilon = 10^{-12}$}

    \addplot [color=mycolor1, mark=triangle]
    table[row sep=crcr, y expr=100*(1-\thisrowno{1})]{%
      15.1181786195608	0.337981586488032\\
      27.5506338369966	0.347631160218381\\
      40.750083606356	0.354260520134288\\
      98.3455584910656	0.393273143877357\\
      342.981909863643	0.499377274500605\\
      516.744465349849	0.541706042328029\\
      784.988124538484	0.587250188243123\\
      1243.73404499377	0.640158280861504\\
      1711.16753188835	0.675171776024926\\
      4593.50842538426	0.77638642693254\\
      6549.24373396621	0.805135230474514\\
      9507.11035234144	0.830967791975928\\
      10497.4824523219	0.841941191557177\\
      14474.0965900823	0.859745947340255\\
    };
    \addlegendentry{$\epsilon = 10^{-15}$}

 \end{axis}
\end{tikzpicture}%

%% file: figures/res_multihop_SNR.tex
\definecolor{darkgreen}{RGB}{0, 128, 0}
\definecolor{mycolor1}{rgb}{1.00000,0.00000,1.00000}%
\begin{tikzpicture}

  \pgfplotsset{
    grid style = {
      dash pattern = on 0.05mm off 1mm,
      line cap = round,
      black,
      line width = 0.5pt
    },
    title style = {font=\fontsize{7pt}{7.2}\selectfont},
    label style = {font=\fontsize{8pt}{7.2}\selectfont},
    tick label style = {font=\fontsize{7pt}{7.2}\selectfont},
  }

  \begin{axis}[%
    width=\columnwidth,
    height=\plotfigureheight\columnwidth,
    grid=major,
    unbounded coords=jump,
    xmin=0.1, xmax=1,
    xtick={0.1,0.2,...,1.1},
    xlabel={Normalized achievable code rate: $R^{*}(\epsilon,P,L)/C$},
    xlabel style={yshift=0.6em},%
    ymin=10, ymax=70,
    ytick={0,10,...,80},
    ylabel={Latency reduction (\%)},
    ylabel style={yshift=-1.6em},
    legend pos={south west},
    legend style={
      anchor={south west},
      cells={anchor=west},
      column sep=0mm,
      row sep=-0.75mm,
      font=\fontsize{7pt}{7.2}\selectfont,
      mark size=2.0pt,
      mark options=solid
    },
    mark size=2.0pt,
    mark options=solid,
    every axis plot/.append style={thick},
    ]
    
    \addplot [color=blue, mark=o]
    table[row sep=crcr, y expr=100*(1-\thisrowno{1})]{%
      -1.15606178128185	nan\\
      -0.937215800223541	nan\\
      -0.774824203295044	nan\\
      -0.387937002922488	nan\\
      0.109284117881763	0.340009121238509\\
      0.243788247197452	0.360425692730713\\
      0.364555750263856	0.401585321659774\\
      0.478880133503872	0.4572288783778\\
      0.54743818598277	0.500180784537995\\
      0.711685438312378	0.636918455459182\\
      0.75580793994622	0.679326050725814\\
      0.79528491366457	0.725091760448287\\
      0.804715728794511	0.73433108560031\\
      0.832519703764389	0.765497788041549\\
      0.854778784788094	0.797027513655046\\
    };
    \addlegendentry{SNR $= -4$ dB}

    \addplot [color=red, mark=x]
    table[row sep=crcr, y expr=100*(1-\thisrowno{1})]{%
      -0.684242086682232	nan\\
      -0.512443861895328	nan\\
      -0.385047961581835	nan\\
      -0.0818720078491214	nan\\
      0.306921412928443	0.374141560668757\\
      0.411899091972448	0.411587811827231\\
      0.50607406319617	0.460635819473507\\
      0.595147936451342	0.518752035070249\\
      0.648525467521028	0.561819485409119\\
      0.776276620521502	0.686488173927676\\
      0.810561771152355	0.728146936749852\\
      0.841224167768315	0.766841894716041\\
      0.848547351046965	0.768382809588961\\
      0.870133226112522	0.806150772555362\\
      0.887409428677614	0.827186386769216\\
    };
    \addlegendentry{SNR $= -2$ dB}

    \addplot [color=darkgreen,mark=square]
    table[row sep=crcr, y expr=100*(1-\thisrowno{1})]{%
      -0.310632229959241	nan\\
      -0.176507846823822	nan\\
      -0.0770922214146609	nan\\
      0.159323446690386	0.344499202952173\\
      0.462069036111191	0.424023174746243\\
      0.543712145808601	0.468452880564855\\
      0.616911618175989	0.519920384136423\\
      0.686106448443125	0.575202686635457\\
      0.727551625913264	0.615615072605322\\
      0.826678375624212	0.731378546008571\\
      0.853264223415043	0.767165438392411\\
      0.877034130365798	0.794975058673018\\
      0.88271018405771	0.807965796318833\\
      0.899438695692262	0.829016523173547\\
      0.912824802193129	0.854069713348953\\
    };
    \addlegendentry{SNR $= 0$ dB}

    \addplot [color=black, mark=diamond]
    table[row sep=crcr, y expr=100*(1-\thisrowno{1})]{%
      0.0942322056054112	0.337329656410074\\
      0.187175926670293	0.338753483412074\\
      0.256042628469436	0.348880350300505\\
      0.419711756517465	0.387955926274481\\
      0.629050546739614	0.494877724231998\\
      0.685445928362409	0.539492497841619\\
      0.735984474133003	0.588034534609822\\
      0.783735182850794	0.642231465881304\\
      0.812324647241941	0.674781294364677\\
      0.880665437463031	0.774358510884175\\
      0.898984493850263	0.80624741807778\\
      0.915359316171859	0.838041733223108\\
      0.919268919535531	0.840733700738838\\
      0.930790018607783	0.858137347822968\\
      0.940007709978255	0.879078363901927\\
    };
    \addlegendentry{SNR $= 3$ dB}

    \addplot [color=mycolor1, mark=+]
    table[row sep=crcr, y expr=100*(1-\thisrowno{1})]{%
      0.281310045623787	0.349807871439509\\
      0.35512194933836	0.360571813856422\\
      0.40980648808702	0.372765917248802\\
      0.539744560715088	0.419014243896541\\
      0.70587569997265	0.526856442917148\\
      0.750616042004896	0.574186360972398\\
      0.790703676880378	0.62090585384673\\
      0.828574051202422	0.671236905556225\\
      0.851244966980092	0.705261637092152\\
      0.905428069695878	0.795252252194264\\
      0.919949514028598	0.826868890659152\\
      0.932928762365751	0.846697523179823\\
      0.936027502154696	0.853657897028587\\
      0.945158749918568	0.875835297267007\\
      0.952464014279563	0.88972305491233\\
    };
    \addlegendentry{SNR $= 5$ dB}

    \addplot [color=orange, mark=triangle]
    table[row sep=crcr, y expr=100*(1-\thisrowno{1})]{%
      0.419101141814707	0.363937240010015\\
      0.478788757794354	0.37897803526629\\
      0.523006419879596	0.394398268795087\\
      0.62806294788433	0.442196387337717\\
      0.762354887542539	0.555841819444482\\
      0.798514377463238	0.597079257796034\\
      0.830910864817924	0.642570589739593\\
      0.861512994484471	0.691148408791135\\
      0.879831562019505	0.722334781188764\\
      0.923608460622776	0.808196123122083\\
      0.935339882316001	0.834270877424357\\
      0.945824986209101	0.859654369603713\\
      0.948328197900534	0.862999597376465\\
      0.955704422610146	0.885832008701105\\
      0.961605458087139	0.895210452161384\\
    };
    \addlegendentry{SNR $= 7$ dB}

  \end{axis}

\end{tikzpicture}%